# Converting Hemodialysis (HD) membranes to Extracorporeal Membrane Oxygenation (ECMO) machines

Abhimanyu Das[a], Matthew L. Robinson[b], D. M. Warsinger[a]

[a] School of Mechanical Engineering and Birck Nanotechnology Center, Purdue University, West Lafayette, IN, 47907, USA
[b] Division of Infectious Diseases, Center for Clinical Global Health Education, Johns Hopkins University School of Medicine, Baltimore, Maryland

April 2020

**ABSTRACT**
Crises like COVID-19 can create a massive unmet demand for rare blood oxygenation membrane machines for impaired lungs: Extracorporeal Membrane Oxygenation (ECMO) machines. Meanwhile, Hemodialysis (HD) machines, which use membranes to supplement failing kidneys, are extremely common and widespread, with orders of magnitude more available than for ECMO machines[1]. This short study examines whether the membranes for HD can be modified for use in blood oxygenation (ECMO). To do so, it considers mass transfer at the micro-scale level, and calculations for the macro-scale level such as blood pumping rates and $O_2$ supply pressure. Overall, while this conversion may technically be possible, poor gas transport likely requires multiple HD membranes for one patient.

**INTRODUCTION**
**Background on HD and ECMO**
HD and ECMO machines both operate by removing large volumes of blood from the body and passing them through membranes that rely on diffusion to transfer desirable or undesirable solutes. HD machines focus on adding and removing compounds to assist with kidney failure, including removing salts and small organic compounds like urea. Meanwhile, ECMO uses membranes to replace $O_2$ and help remove $CO_2$.

**Conversion approach requirements for HD for ECMO**
Custom HD machines made for blood oxygenation[2] have been successfully demonstrated in the past; the question is whether existing HD machines can be converted readily. The blood flow paths of the two machines are similar, although flow rates and pipe diameters are smaller in HD: this may require multiple machines. Catheter insertion points for ECMO are chosen to provide $O_2$ where it is needed; the same placement would be necessary for a HD to ECMO conversion. The dialysate side of the HD membrane presents numerous problems that limit the use of a liquid dialysate. While it is possible to use an oxygenated dialysate for ECMO, it would need a separate oxygenation chamber along with high pressure $O_2$ supply. On the contrary, directly connecting a $O_2$ supply to the dialyzer ports allows for a far simpler arrangement that requires a lower $O_2$ supply pressure. Therefore, it was assumed that a pressurized $O_2$ gas would be far more viable, although it presents its own concerns like bypassing the machine warning and shut-down systems triggered by no flow through the dialysate pump. Such pressurized $O_2$ tanks are common in hospitals. Simply described, such a conversion would resemble the invasive setup of venovenous ECMO (VV-ECMO), with blood connected to HD machine(s) that house the membranes, pump and controls, and the dialysate side connected to a high pressure $O_2$ tank.

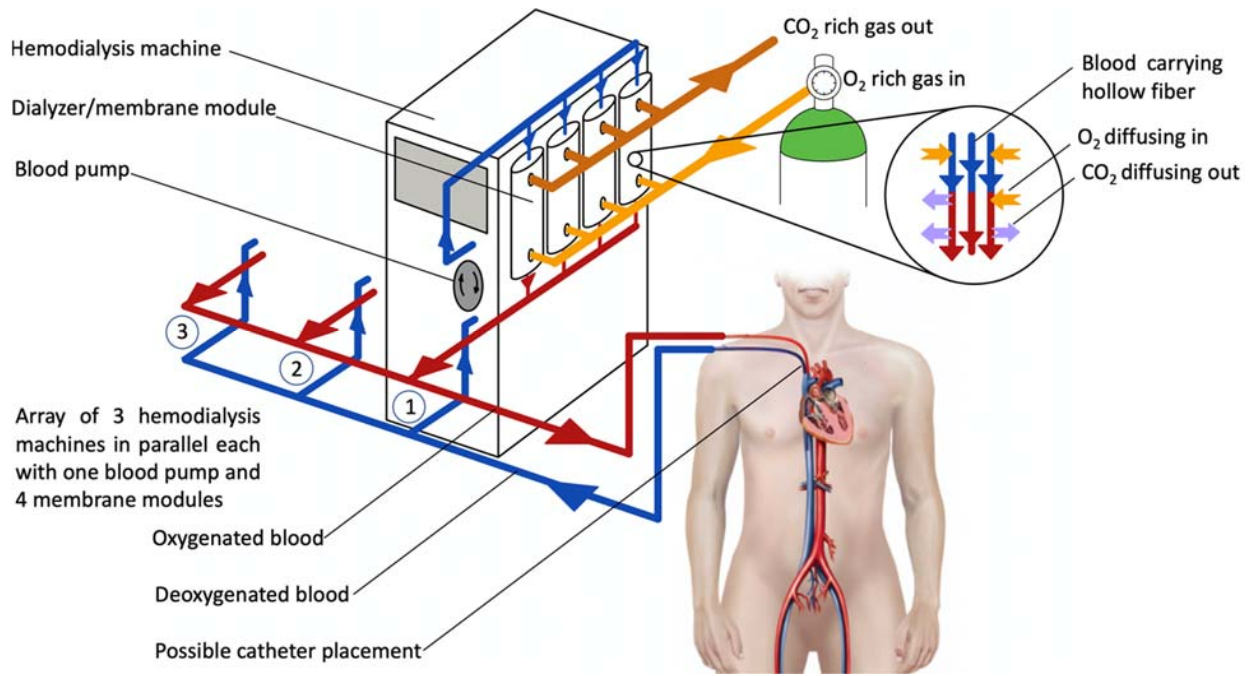

**Figure 1.** Schematic of HD machines converted to operate as a VV-ECMO machine

The number of HD machines required to be connected in parallel for the conversion depends on the maximum flow rate through the blood pump. With a flow rate of 1.63 l min$^{-1}$ through the blood pump, an array of at least three HD machines is needed for the conversion. The number of membrane modules required depends on the blood flow rate through them. While the blood circuit is the same as in a conventional dialysis machine, the dialyzer circuit is modified to operate with $O_2$ gas from a compressed cylinder or a medical gas supply line. Gas exchange occurs inside the dialyzer as deoxygenated blood gets oxygenated while rejecting $CO_2$ to the gas stream due to partial pressure differences of $O_2$ and $CO_2$ across a hollow fiber. Modified human body graphic from http://www.learnecmo.com/vv-ecmo

**Concerns**

There are numerous concerns and risks to address for attempting to modify HD machines as ECMO machines.

- Reliability: putting a system together with multiple machines and membranes, calculated here as 3 HD machines and 12 membranes modules
- Each patient takes up at least 3 dialysis machines
- High $O_2$ supply pressure may lead to bubble induction in blood
- Pumps in parallel may lead to flow instabilities
- Peristaltic blood pumps in HD machines may have a maximum flow rate less than 1.6 l min$^{-1}$
- Modifications to dialysis machines may be difficult to make
- Continuous renal replacement therapy (CRRT) systems might be needed as these systems will be used for long periods of time as compared to typical dialysis machines that are used for a few hours

- Calculations have been carried out using values from multiple sources and topics; actual membrane properties of commercially available membranes have not been used
- Mass transfer resistance on blood side has been neglected; actual membrane gas permeability may be lower than calculated

# NOMENCLATURE

| Symbol | Meaning | Units |
|---|---|---|
| $d_i$ | Internal diameter of hollow fiber | μm |
| $t$ | Wall thickness of hollow fiber | μm |
| $d_o$ | External diameter of hollow fiber | μm |
| $L$ | Length of hollow fiber | cm |
| $A$ | Membrane area | m² |
| $N_f$ | Number of hollow fibers in a membrane module | [-] |
| $p_{ven}$ | Partial pressure of $O_2$ in systemic venous blood | mm Hg |
| $p_{art}$ | Partial pressure of $O_2$ in systemic arterial blood | mm Hg |
| $f_{ven}$ | Volume fraction of $O_2$ in systemic venous blood | [-] |
| $f_{art}$ | Volume fraction of $O_2$ in systemic arterial blood | [-] |
| $D_{O_2,plasma}$ | Bulk diffusivity in $O_2$ in plasma | cm² s⁻¹ |
| $\varepsilon$ | Porosity of membrane material | [-] |
| $\tau$ | Tortuosity of membrane material | [-] |
| $D_{eff}$ | Effective diffusivity of $O_2$ in membrane material | cm² s⁻¹ |
| $\alpha$ | Solubility of $O_2$ in plasma | ml $O_2$ ml plasma⁻¹ mm Hg⁻¹ |
| $K$ | Krogh's diffusivity of $O_2$ in membrane material | cm² s⁻¹ ml $O_2$ ml plasma⁻¹ mm Hg⁻¹ |
| $k_m$ | Mass transfer coefficient of hollow fiber wall | ml min⁻¹ mm Hg⁻¹ |
| $LMPD$ | Logarithmic mean pressure difference | mm Hg |
| $\dot{Q}_{supply}$ | Rate of blood oxygenation in a membrane module | ml min⁻¹ |
| $\dot{Q}_{cons}$ | Rate of $O_2$ consumption in human body | ml min⁻¹ |
| $\dot{Q}_{b,total}$ | Total blood flow rate from the body | ml min⁻¹ |
| $\dot{Q}_{b,d,max}$ | Maximum flow rate through HD blood pump | ml min⁻¹ |
| $\dot{Q}_{b,rated}$ | Rated blood flow rate through a membrane module | ml min⁻¹ |
| $N_d$ | Number of HD machines required in parallel | [-] |
| $\dot{Q}_{b,d}$ | Actual flow rate through HD blood pump | ml min⁻¹ |
| $N_m$ | Number of membrane modules required in parallel | [-] |
| $\dot{Q}_{b,m}$ | Actual blood flow rate through a membrane module | ml min⁻¹ |
| $\dot{Q}_{demand}$ | Rate of blood oxygenation in a membrane module | ml min⁻¹ |
| $p_{O_2}$ | Partial pressure of $O_2$ on gas side | kPa |
| $f$ | Volume fraction of $O_2$ in gas supply | [-] |
| $p_{tot}$ | Total pressure of gas supply | kPa |
| $\eta$ | Viscosity of whole blood | Pa s |
| $\Delta p_b$ | Pressure drop along a hollow fiber | kPa |

# METHODOLOGY

The approach analyzes blood flow rates for $O_2$ supply and performs simple transport calculations to examine whether sufficient gas exchange can be provided. The concentrations of $O_2$ present are described in partial pressures. Representative values of HD membrane properties were obtained.

A large number of commercially available synthetic HD membranes today are made of polysulfone[3] which is inherently hydrophobic. Hydrophilicity is imparted in the membrane material by adding polyvinylpyrrolidone (PVP) for reasons stated in [4]. Each membrane module consists of tens of thousands of hollow fibers through which blood flows while the dialysate flows over the fibers[4]. The two flows are usually counter-current, with radial concentration gradients driving mass transfer of various necessary and waste solutes. This arrangement differs from ECMO in which blood flows over the hollow fibers which carry oxygen[5]. We adhere to the former arrangement with $O_2$ flowing on the dialysate side to keep the number of necessary modifications at a minimum.

The membrane dimensions used in the calculations are listed in Table 1. For properties that have been reported with a range, average values were used. The value for membrane area was chosen from the largest available membrane from Fresenius Medical Care to assess a system with least number of membrane modules.

The external radius of the hollow fiber calculated as

$$d_o = d_i + 2t \qquad (1)$$

while the number of fibers in a membrane module were calculated using

$$N_f = \frac{A}{\pi d_o L} \qquad (2)$$

**Table 1**

| Membrane properties | Values |
|---|---|
| $d_i$ (internal diameter) | 180-220 μm[4] |
| $t$ (wall thickness) | 20-50 μm[6] |
| $d_o$ (external diameter) | 270 μm |
| $L$ (length of hollow fiber) | 20-24 cm[6] |
| $A$ (total area) | 2.5 m²[3] |
| $N_f$ (number of fibers) | 13397 |

The transport of gases between the blood and the oxygen streams, across the membrane, is essentially a diffusion problem driven by concentration gradients. Since $O_2$ and $CO_2$ are carried by the blood as species bound to hemoglobin, we need to the convert the concentration gradients to partial pressure gradients in order to be able to use the oxygen dissociation curve (ODC). The ODC is a non-linear relationship between the volume fraction of $O_2$ in blood and its partial pressure.

Table 2 lists the physiological parameters used for calculating the required $O_2$ transport, indicating the operating points of the ODC curve. The values used are representative of a healthy adult and are the target values that need to be achieved through ECMO in a patient.

**Table 2**

| Physiological parameters | Values |
|---|---|
| $p_{ven}$ (Partial pressure of $O_2$ in systemic venous blood) | 40 mm Hg[7] |
| $p_{art}$ (Partial pressure of $O_2$ in systemic arterial blood) | 95 mm Hg[7] |
| $f_{ven}$ (Volume fraction of $O_2$ in systemic venous blood) | 0.157[8] |

| $f_{art}$ (Volume fraction of O₂ in systemic arterial blood) | 0.208[8] |

In order to solve for the rate of diffusion of O₂, the diffusion coefficient or diffusivity of O₂ in the membrane material needs to be calculated. Given the partly hydrophilic nature of the membrane material, we assume that the pores in the hollow fibers are mostly filled with plasma. The diffusivity of O₂ in pore confined plasma, however, would be lower than its bulk value, ($D_{O_2,plasma}$). Given the porosity ($\varepsilon$) and the tortuousness of the transport path ($\tau$) in a polymer material, an effective diffusivity can be calculated as

$$D_{eff} = \frac{\varepsilon D_{O_2,plasma}}{\tau} \quad [9] \qquad (3)$$

The effective diffusivity can then be multiplied with the solubility of O₂ in plasma ($\alpha$) to obtain Krogh's diffusivity given as

$$K = \alpha D_{eff} \quad [10] \qquad (4)$$

This along with the partial pressures of O₂ allows us to use the same functional forms of the diffusion equations to calculate O₂ transport across the membrane. Reported and calculated values of the diffusion parameters are listed in Table 3.

**Table 3**

| Diffusion parameters | Values |
| --- | --- |
| $D_{O_2,plasma}$ (at 37°C) | 2.18e-5 cm² s⁻¹[11] |
| $\varepsilon$ | 0.3[12] |
| $\tau$ | 3.33[13] |
| $D_{eff}$ | 1.98e-6 cm² s⁻¹ |
| $\alpha$ | 0.003 ml O₂ (100 ml plasma)⁻¹ m Hg⁻¹[10] |
| $K$ | 5.89e-11 cm² s⁻¹ ml O₂ ml plasma⁻¹ mm Hg⁻¹ |

**Transport calculations**

The transport calculations must consider the micro-scale, describing diffusion through membranes into the blood, and the macro-scale, describing changing O₂ levels with length. Relative diffusion in the different channels must be compared to neglect those that are an order of magnitude smaller.

On the micro-scale, O₂ permeation can be calculated by assuming a negligible gas side resistance and ignoring the blood side resistance owing to a high blood flow rate[13]. With mass transfer being dominated by the membrane resistance, the overall mass transfer coefficient for a single hollow fiber can be written as

$$k_m = \frac{2\pi LK}{\ln(\frac{d_o}{d_i})} \qquad (5)$$

On the macro-scale, the variation of gas concentration with length along the membranes resembles heat transfer in a shell and tube heat exchanger with phase change on the shell side, so a logarithmic mean pressure difference is calculated as

$$LMPD = \left[\frac{(p_{O_2}-p_{ven})-(p_{O_2}-p_{art})}{\ln\left(\frac{p_{O_2}-p_{ven}}{p_{O_2}-p_{art}}\right)}\right] \quad (6)$$

The partial pressure of $O_2$ on the gas side and is assumed to be constant along the length of a hollow fiber. This appears to be a reasonable assumption provided the rate of $O_2$ transport across the membrane is small as compared to the bulk $O_2$ flow rate on the gas side.

The net rate of blood oxygenation in one membrane module with $N_f$ fibers can then be written as

$$\dot{Q}_{supply} = N_f k_m LMPD \quad (7)$$

The blood flow rate from the body necessary to satisfy its oxygen demand ($\dot{Q}_{cons}$) can be calculated as

$$\dot{Q}_{b,total} = \frac{\dot{Q}_{cons}}{f_{art}-f_{ven}} \quad (8)$$

The flow rates listed in Table 4 can be used to calculate the required $O_2$ supply pressure and the number of HD machines and membranes modules.

**Table 4**

| Flow rates | Values |
| --- | --- |
| $\dot{Q}_{cons}$ (Rate of $O_2$ consumption in human body) | 250 ml min$^{-1}$[7] |
| $\dot{Q}_{b,total}$ (Total blood flow rate from the body) | 4902 ml min$^{-1}$ |
| $\dot{Q}_{b,d,max}$ (Maximum flow rate through HD blood pump) | 2000 ml min$^{-1}$ |
| $\dot{Q}_{b,rated}$ (Rated blood flow rate through a membrane module) | 400 ml min$^{-1}$[3] |

The number of HD blood pumps needed is obtained using

$$N_d = ceiling\left(\frac{\dot{Q}_{b,total}}{\dot{Q}_{b,d,max}}\right) \quad (9)$$

The flow rate through each HD blood pump is then

$$\dot{Q}_{b,d} = \frac{\dot{Q}_{b,total}}{N_d} \quad (10)$$

Similarly, the total number of membranes required in parallel

$$N_m = N_d floor\left(\frac{\dot{Q}_{b,total}}{N_d \dot{Q}_{b,rated}}\right) \quad (11)$$

The above expression for $N_m$ ensures that the blood flow rate through each membrane module is higher than the rated value and that each dialysis machine handles the same number of membranes allowing for identical arrangement.

The actual blood flow rate through each membrane module is then

$$\dot{Q}_{b,m} = \frac{\dot{Q}_{b,total}}{N_m} \quad (12)$$

The required rate of oxygenation in each bundle is then calculated as

$$\dot{Q}_{demand} = \dot{Q}_{b,m}(f_{art} - f_{ven}) \quad (13)$$

The required partial pressure of O₂ on gas side is obtained ($p_{O_2}$) by equating $\dot{Q}_{supply}$ with $\dot{Q}_{demand}$; the calculated values have been listed in Table 5 along with other parameters. The value obtained is less than the usual maximum supply pressure of O₂ in hospitals[14].

The calculated value of $p_{O_2}$ can be achieved by either using pure O₂ at the same pressure or a mixture with lower volume fraction ($f$) and higher total pressure ($p_{tot}$) according to

$$p_{O_2} = f p_{tot} \quad (14)$$

Assuming blood to be a Newtonian fluid[4], [15], the pressure drop along the hollow fiber is calculated as

$$\Delta p_b = \frac{128 \eta L \dot{Q}_{b,m}}{N_f \pi d_i^4} \quad (15)$$

In equation above, the viscosity of whole blood ($\eta$) used instead of that of plasma to estimate maximum pressure drop.

**Table 5**

| Parameter | Value |
| --- | --- |
| $k_m$ | 1.63e-6 ml min⁻¹ mm Hg⁻¹ |
| $N_d$ | 3 |
| $\dot{Q}_{b,d}$ | 1634 ml min⁻¹ |
| $N_m$ | 12 |
| $\dot{Q}_{b,m}$ | 408.5 ml min⁻¹ |
| $\dot{Q}_{demand}$ | 20.83 ml min⁻¹ |
| LMPD | 956.2 mm Hg |
| $p_{O_2}$ | 134.74 kPa |
| $\eta$ | 4e-3 Pa s[7] |
| $\Delta p_b$ | 11.4 kPa |

## CONCLUSION

The limiting factor in the above arrangement is the maximum flow rate through the HD blood pump, which in turn dictates the number of dialysis machines required to be put in parallel (3 here). If the maximum blood pump flow rate is lower than 1.63 l min$^{-1}$, we will need another dialysis machine and so on. Another major concern would be to reprogram a hard-coded HD console that might be not function as desired in the absence of the dialysate loop. Overall, there a significant challenges in gas exchange in this conversion, but these likely can be overcome with engineering to use multiple membranes per patient, modified flow and catheters, and the inclusion of high pressure $O_2$.

## ACKNOWLEDGEMENTS
We would like to thank Maher Amadi, William Clark, Samir C Gautam, and Steve Sozio for their feedback on this work.